\documentclass[prd,aps,twocolumn,floatfix]{revtex4}
\usepackage{epsf}
\newcommand{\be}{\begin{equation}}
\newcommand{\ee}{\end{equation}}
\newcommand{\ba}{\begin{eqnarray}}
\newcommand{\ea}{\end{eqnarray}}
\newcommand{\eb}{\end{thebibliography}}

\newcommand{\bb}{{\mathbf b}}

\begin{document}
\title{New analytic unitarization schemes }
\author{J.-R. Cudell\footnote{J.R.Cudell@ulg.ac.be},
E. Predazzi\footnote{predazzi@to.infn.it} and O.V. Selyugin \footnote{selugin@theor.jinr.ru}}
\affiliation{
$^*$ IFPA, AGO Dept., Universit\'e de Li\`ege, Li\`ege, Belgium\\
$^\dagger$ \it Dipartamento di Fisica Teorica, Universit\`a di Torino, Italy\\
$^\ddagger$ \it BLTP,
Joint Institute for Nuclear Research,
Dubna, Russia}

\pacs{...}


\begin{abstract}
  We consider two well-known classes of unitarization of Born amplitudes
  of hadron elastic scattering.
  The standard class, which
  saturates at the black disk limit  includes the
  standard eikonal representation, while the other class, which goes beyond the
  black disk limit to reach the full unitarity circle,
  includes the U matrix.
 It is shown that the basic  properties of these schemes
  are independent of the functional form used for the unitarisation, and that
  U-matrix and eikonal schemes can be extended to have similar properties.
  A common form of  unitarization is proposed  interpolating between both classes.
  The correspondence with different nonlinear equations are also briefly  examined.
\end{abstract}

\maketitle

\section{Introduction}
At low energies, hadron scattering can be described by one-Reggeon exchange
terms.
But as the Pomeron term(s) grow(s) with energy, these exchanges will
eventually
violate unitarity. To see this, one can switch to partial waves
\cite{Predazzi66}, or to the representation in impact parameter $\bb$. As
$s$ grows, one needs to sum many partial waves with
 $l ~\sim|\bb|\sqrt{s}\rightarrow \infty $. The summation over $l$ then
becomes an integration over $\bb$.

The partial wave $G(s,\bb)$ has two regimes. First of all, it can reach
maximum
inelasticity. In this case, $G(s,\bb)=1$, and half of the interactions are
inelastic.
The center of the protons then becomes black, and multiple exchanges, {\it
i.e.}
cuts in the complex $J$ plane, become important. Second, the partial
wave can
 later reach the full unitarity limit $G(s,\bb)=2$.

The maximum inelasticity limit may be reached
in $pp$ or $p\bar{p}$ scattering
a little above the Tevatron energy
\cite{dif04}, so that one expects cuts to be important in the description
of soft
interactions at the CERN  LHC. The inclusion of these goes under the name of
unitarization.
It is a formidable task to calculate the contribution of cuts, as not
only multiple
Pomeron exchanges must be calculated, but also multiple Pomeron vertices.

Different schemes have been proposed, and we want in this paper to show
that
the general properties of the amplitude do not heavily depend on the
scheme,
but rather on what assumes for the inelastic contribution at high energy.
We shall
limit ourselves to two popular schemes: the eikonal and the $U$ matrix,
and
show that simple extensions of each lead to similar properties.

In Sec. I, we remind the reader of the simple requirements coming from
unitarity,
and examine in Sec. II the two schemes. In Sec. III, we show that it is
possible
to obtain the properties of the eikonal by extending the $U$-matrix scheme,
whereas in
Sec. IV, we show the reverse, {\it i.e.} that one can extend the eikonal
to mimic the behavior of the $U$ matrix, at least for physical amplitudes.

\section{Unitarity}
At high energy, we can start with the elastic scattering amplitude
$a(s,t)$, related to
the elastic cross section though
\begin{equation}
\frac{d\sigma}{dt}=\frac{1}{16\pi s^2} |a(s,t)|^2.
\end{equation}
One can then Fourier transform $a$ to $\bb$ space
\begin{equation}
G(s,\bb)=\int\frac{d^2{\mathbf\Delta}}{(2\pi)^2}\frac{a(s,t)}{2s}e^{i\bf\Delta\cdot\bb},
\end{equation}
which leads to the expressions
\begin{eqnarray}
\sigma_{tot}&=&2\int d^2\bb\ \mathrm{Im}\ G(s,\bb),\\
\sigma_{el}&=&\int d^2 \bb\ \left|G(s,\bb)\right|^2,
\end{eqnarray}
where we have assumed that the spin-flip contribution to the elastic cross
section is negligible.
One can then write the square of the $S$-matrix density $S(s,\bb)=1+i G(s,
\bb)$ as
\begin{equation}
|S(s,\bb)|^2=1-2\mathrm{Im}\ G(s,\bb)+\left|G(s,\bb)\right|^2.
\end{equation}
Unitarity demands that $|S(\bb)|^2|\leq 1$, the difference coming
from inelastic channels:
\be
\eta_{in}(s,\bb)=1-|S(s,\bb)|^2\geq 0.
\ee

There are several ways to represent the unit circle. First of
all, one can map the upper complex plane into a circle via a complex exponential
\be
S(s,\bb)=\exp(i z(s,\bb)) \ {\rm with}\ {\rm Im} z(s,\bb)\geq 0,\ .
\label{erep}\ee
This maps in fact an infinite number of strips with $2n\pi < {\rm Re} z(s,\bb) < 2(n+1)\pi$
each onto the unit circle.

It is also possible to use a one-to-one map through a M\"obius transform
and write
\be
S(s,\bb)=\frac{1+i z'(s,\bb)}{1-i z'(s,\bb)},  \ {\rm with}\ {\rm Im} z'(s,
\bb)\geq 0.
\label{urep}\ee
Other representations are possible,
but we shall concentrate on these two  in the following section.

\section{Unitarization}
The physical amplitude lies within the unitarity circle, so that the
associated $S$
matrix can always be represented by Eqs.~(\ref{erep}) and (\ref{urep}).
The unitarization scheme comes in once one identifies $z$ or $z'$ with the
one-Reggeon
exchange amplitude. One then considers (\ref{erep}) and (\ref{urep}) as
series
expansions in n-Reggeon exchanges, so that their first term must give
$1+i\chi(s,b)$.

Indeed, if one writes the one-Reggeon exchange amplitude as $\chi(s,\bb)$,
then
assuming $z=\chi$  in (\ref{erep}) leads to the well-known
eikonal
representation:
\be
G(s,\bb)=i(1-\exp(i\chi(s,\bb)).
\ee
This scheme can be derived in QED and other field theories
\cite{Shiff,ChengWu,Blokhintzev,Barbashev} or in potential theory
\cite{Goldberger}.
It can be extended to include diffractive channels \cite{Gotsman}.
It leads at  asymptotic energies ($s
\rightarrow
 \infty$) to  the limit $\sigma_{{\rm el}}/\sigma_{{\rm inel}} = 1$,  i.e. to
maximum inelasticity.

The other unitarization scheme considered here is the $U$-matrix
representation
\cite{bg,Logunov,Chrustalev} where one identifies $z'$ in (\ref{urep})
with $\chi(s,\bb)/2$, to match the one-Reggeon
exchange
\begin{eqnarray}
 G(s,\bb)  =  \frac{ {\chi}(s,\bb)}{1 \ -i {\chi(s,\bb)}/2}.
\label{U-mt}
\end{eqnarray}
 In this scheme,  $S(s,\bb)$ tends to $-1$ when $s\rightarrow \infty$
 and $\bb$ is finite, so
that the inelastic partial wave $\eta_{{\rm in}}(s,\bb)$ tends to $0$: the ratio
$\sigma_{{\rm el}}/\sigma_{{\rm inel}}$ vanishes asymptotically.

Both schemes have the same development at second order in $\chi$, and
differ only
in the rest of the series.

It must be noted however that the resummation must lead to an amplitude
within the
unitarity circle, but there is no reason to assume that it maps the entire
complex plane to
the circle. Hence, one can easily extend both schemes through a change in
the
strength of successive scattering. This gives the extended eikonal
schemes
\cite{Martiros,Kaidalov1,Martynov1}
\be
G(s,\bb)=\frac{i}{\omega}(1-\exp(i\omega\chi(s,\bb))\label{exteik}
\ee
and the extended $U$-matrix schemes
\begin{eqnarray}
 G(s,\bb)  =  \frac{ {\chi}(s,\bb)}{1 \ -i {\omega'\chi(s,\bb)}}.
\label{extumat}
\end{eqnarray}
It is straightforward to check that using $\omega \geq 1$ or $\omega^{\prime} \geq 1/2$
maps any amplitude $\chi$ into the unitarity circle.

We shall now show that the various possibilities can be grouped
into two
wide classes of unitarization schemes, and that the exact form matters
little.
\section{Shadowing}
As we have seen, the eikonal predicts that at high energy the inelastic
component of
the cross section will be maximal, $\eta_{{\rm in}}=1$. This in turn leads to $|
S(\infty,\bb)|
=0$ and $G(\infty,\bb)=i$. To reach this regime via an extended $U$
matrix, one
needs to choose $\omega'=1$ in (\ref{extumat}).

The inelasticity  will then be
\begin{eqnarray}
\eta_{{\rm in}}(s,\bb) \ = \  \frac{ 2{\rm Im}\chi(s,\bb) +  |\chi(s,\bb)|^2}
{1+2{\rm Im}\chi(s,\bb)+|
\chi(s,\bb)|^2} \nonumber \\ .
\end{eqnarray}
  It can easily be seen that  $ s \rightarrow  \infty$
leads to
  $\sigma_{{\rm inel}}/\sigma_{{\rm el}} \rightarrow \ 1 \ $ and
   $\sigma_{{\rm el}}/\sigma_{{\rm tot}} \rightarrow \ 1/2$.

Differently stated, this  extended  $U$ matrix representation has the
standard
black disk limit.

\begin{figure}[!ht]
\epsfysize=45mm
\epsfbox{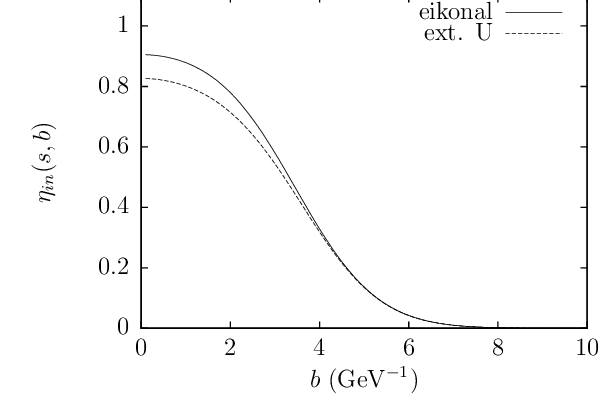}
\epsfysize=45mm
\epsfbox{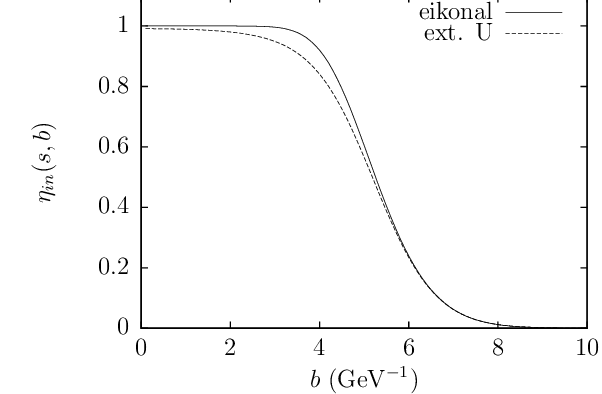}
\caption{Inelasticity for the eikonal and
 extended $U$ matrix at $\sqrt{s}=1.8 \ $TeV (upper figure) and
     $\sqrt{s}=14 \ $TeV  (lower figure).  }
\end{figure}

Throughout this paper, we shall use as an example of one-Reggeon exchange
amplitude
a hard Pomeron term, with a parametrization
\begin{eqnarray}
\chi(s,b)&=&\left(\frac{\sqrt{s}}{1500\ {\rm GeV}}\right)^{0.9}
exp(-\bb^2/(9 \ GeV^{-2})\nonumber\\
&\times&\left((i+\tan\left(\frac{0.45\pi}{2}\right)\right).
\label{hardpom}
\end{eqnarray}
This amplitude reaches the black disk limit at $1500 \ $ GeV, similarly to the
model
of ref.~\cite{physlett}, and has a dependence in $t$ similar to that of
$pp$ scattering. We also neglect the effect of shrinkage, which is small
for a hard Pomeron.

In Fig. 1, one can compare the inelasticity
in the case of the eikonal and in that of the extended $U$ matrix.
One clearly sees that the generic features of both schemes are very close.

The extended $U$ matrix and the eikonal scheme are
different representations of a wider class of unitarization procedures with  a
standard black disk limit. Indeed, we can extend (\ref{extumat}) to
\begin{eqnarray}
  G(s,\bb)  = i[1 \ - \ \frac{1}{(1 \ - \ i\chi(s,\bb)/\gamma)^\gamma}].
\label{2cl}
\end{eqnarray}
If $\gamma=1$ this form leads to the  extended $U$ matrix while, for
$\gamma \rightarrow \infty $,
we obtain the standard  eikonal. When $\gamma$ varies from $1$ to
  $\infty$, we obtain different forms of  unitarization which all lead to
a black disk limit, and the amplitude $G$ does not change anymore once it
has reached its maximum value.
In  Fig. 2, we show the inelasticity $\eta(s,\bb)$ for different
values of $\gamma$ , again in the case of the hard Pomeron input of
Eq.~(\ref{hardpom}).
%
\begin{figure}[!ht]
\epsfysize=45mm
\epsfbox{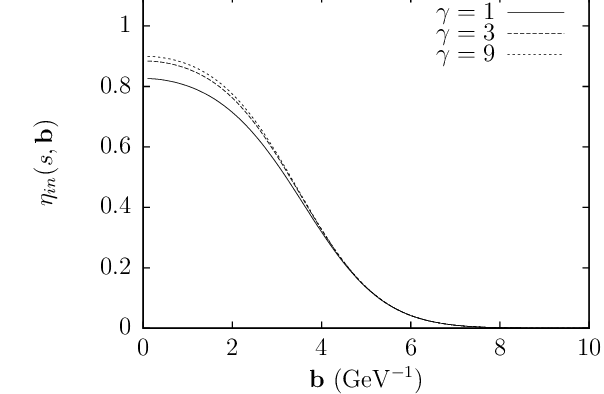}
\epsfysize=45mm
\epsfbox{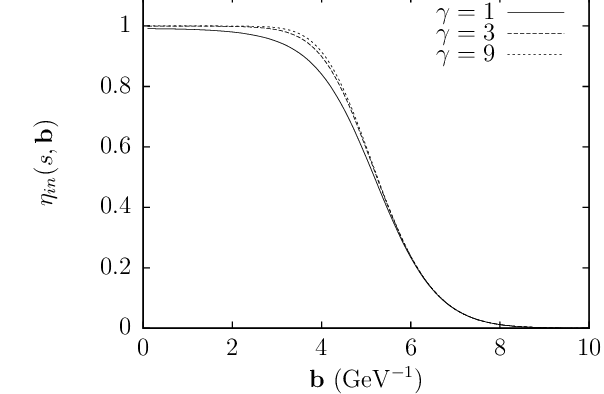}
\caption{Inelasticity $\eta_{{\rm in}}(s,\bb)$ for various values of $\gamma$,
  for  $\sqrt{s}=1.8 \ $TeV (upper line) and
     $\sqrt{s}=14 \ $TeV  (lower line). }
\end{figure}

It may be worth pointing out  that Eq.~(\ref{2cl}) can also lead to
a noninteracting theory in the limit $n\rightarrow 0$, as in this case $G(s,
\bb)\rightarrow 0$.
\section{Full unitarity limit}
The standard $U$-matrix scheme (\ref{U-mt}) was intensively  explored in
\cite{Chrustalev} in the partial-wave language.
In the impact parameter  representation,  the properties of the $U$-matrix
are  explored in \cite{TT1}.

For a purely imaginary one-Reggeon exchange, when ${\rm Im} \chi(s,\bb)$ goes
from $0$ to $ \infty$,
the $S$-matrix varies in the interval $[-1,+1]$, and the amplitude
$G$ goes from 0 to $2i$. Hence in this case the full unitarity limit can
be reached.

This form of  unitarization leads to  unusual properties at super-high
energies as was shown in \cite{TT1}.
In this representation
\begin{eqnarray}
 \sigma_{{\rm el}}(s) \ = \ 4   \ \int_{0}^{\infty}
        \frac{\left|\frac{\chi(s,\bb)}{2}\right|^2}{\left|1 \ -i \left(\frac{\chi(s,\bb)}{2}\right)\right|^2}  \ d\bb
\label{umel1}
\end{eqnarray}
and
\begin{eqnarray}
 \sigma_{{\rm inel}}(s)  \ = \ 4   \ \int_{0}^{\infty}
        {\rm Im} \left(\frac{\left(\frac{\chi(s,\bb)}{2}\right)}{\left[1 \ -i \left(\frac{\chi(s,\bb)}{2}\right)\right]^2}\right)  \ d\bb
        \label{uminel1}.
\end{eqnarray}
so that, when ${\rm Im } \chi(s,\bb)\rightarrow\infty$, one gets
$\sigma_{{\rm inel}}/\sigma_{{\rm tot}}\rightarrow 0$ and
$ \sigma_{{\rm el}}/\sigma_{{\rm tot}} \rightarrow 1$.

It is often considered  \cite{TT07} that these properties are intrinsic to
the M\"obius projection of $\chi$ onto the unitarity circle. We want to
show now
that, in fact, extended eikonals can lead to the same properties for the
unitarized amplitude.

We have seen that choosing $\omega>1$  in (\ref{exteik}) guarantees that
any amplitude would be unitarized. However, one must be concerned with
the physical amplitude, and it is not needed to map the whole complex
plane into the unitarity circle. This means that for some specific choices
of one-reggeon exchange
$\chi$, one can extend the range of values of $\omega$, and restrict
oneself to
part of the complex $\chi$ plane. Unitarity, in this case, leads to the
condition
\begin{equation}
\cos(\omega {\rm Re} \chi)\geq \frac{e^{-\omega {\rm Im } \chi}-(2\omega-1)
e^{\omega {\rm Im }\chi}}{2(1-\omega)}.
\end{equation}
So we see that for $\omega<1/2$, the second term of the numerator will
guarantee the inequality for sufficiently large ${\rm Im } \chi$.

\begin{figure}
\epsfysize=45mm
\epsfbox{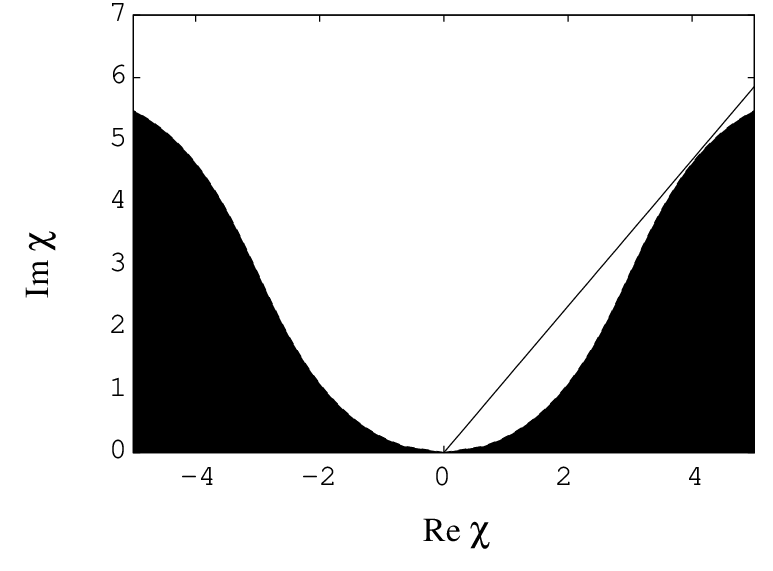}
\caption{The allowed region (in white) for amplitudes $\chi$ to be
unitarised by an
extended eikonal with $\omega=0.525$, together with the line corresponding
to a
hard pomeron amplitude with intercept 1.45. \label{omega}}
\end{figure}

We show in Fig.~\ref{omega} the region allowed in the case $\omega=0.525$
together with a curve showing the amplitude corresponding to the exchange
of one hard pomeron with intercept 1.45. This is of course an extreme
curve, corresponding to a ratio ${\rm Re } \chi/{\rm Im } \chi$ of 0.73. Any
physical amplitude will include
softer intercepts, and will lie above the hard pomeron line. So we see
that, in
practice, eikonals can be extended to values of $\omega$ between 1/2 and
1.

But  at high energy, such eikonals have all the basic properties of the
$U$-matrix unitarization. For instance, the inelasticity reaches the
asymptotic value
\begin{equation}
\eta_{in}\rightarrow \frac{2\omega-1}{\omega^2} {\rm \ as\ }
s\rightarrow\infty
\end{equation}
which is close to 0 for $\omega$ close to $1/2$.

Our calculations
for $\eta_{{\rm in}}(s,b)$ in the cases of the $U$-matrix and of the extended
eikonal
are shown in Fig. 4.
 We see that  both solutions have the same behavior in $s$ and
 $b$,  but also that the extended eikonal has  sharper anti-shadowing
properties.

\begin{figure}[!ht]
\epsfysize=45mm
\epsfbox{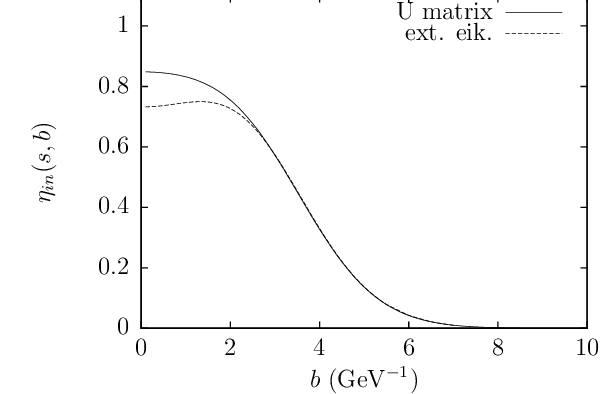}
\epsfysize=45mm
\epsfbox{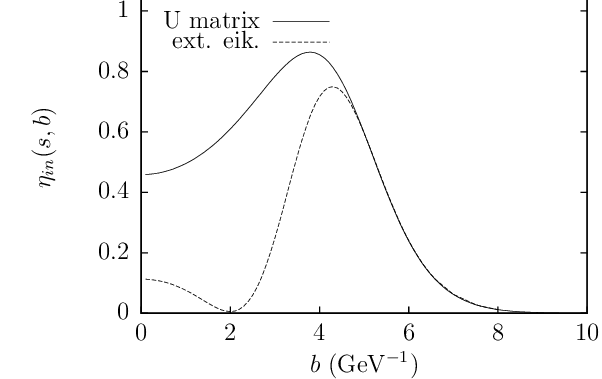}
\caption{ Antishadowing effects in extended eikonal  and
 $U$ matrix schemes, for  $\sqrt{s}=1.8 \ $TeV (upper figure) and
     $\sqrt{s}=14 \ $TeV  (lower figure).
}
\end{figure}

Hence, the peculiar asymptotic properties of
(\ref{uminel1}) are  not unique to the $U$-matrix
(\ref{U-mt}).
The extended eikonal  (\ref{exteik}) has a similar
asymptotic behavior for $\omega$ between 1/2 and 1.

Again, we can find a scheme that interpolates between these two forms
\cite{Miettinen}:
\begin{equation}
G(s,b)  = \frac{i}{\omega}[1  -  \frac{1}{(1  -  i\omega{\chi}
(s,b)/\gamma)^\gamma}].
\label{1cl}
\end{equation}
If $\gamma=1$ this form coincides with the standard $U$-matrix representation
for $\omega=1/2$.
If, on the contrary, we let $\gamma \rightarrow \infty $, we recover the
extended form of the eikonal representation. Hence, when $\gamma$ varies
from $1$ to
$\infty$, and $\omega\approx 1/2$, we obtain  different forms of
unitarization belonging to a
wide class with the similar asymptotic properties, which we shall
examine more closely in the next section.
\section{Comparison of the Born terms}
Another way to make the unitarization schemes lead to similar results
is to use different inputs for $z$ in (\ref{urep}) or (\ref{erep}).
We can solve the equation
\begin{equation}
\frac{i}{\omega}(1-\exp(i\omega\chi_e)
=  \frac{ {\chi_u}}{1 \ -i {\omega'\chi_u}}.
\end{equation}
to find which Born term in the U-matrix representation would give
results similar to those of the eikonal. Writing $\chi_u=\chi_u^R
+i\chi_u^I$, and $\chi_e=\chi_e^R
+i\chi_e^I$, we obtain
\be 2\omega\chi_e^I=\log\left(
|\chi_u|^2 \omega'^2 + 2 {\chi_u^I} \omega' + 1
\over
|\chi_u|^2 (\omega - \omega')^2
- 2 {\chi_u^I} (\omega-\omega') + 1\right)\ee

\be \tan(\omega\chi_e^R)=
{ \chi_u^R \omega
\over
|\chi_u|^2 \omega'(\omega'-\omega) + {\chi_u^I} (2\omega'-\omega) + 1}\ee

This simplifies to a particularly simple expression in the case of a purely imaginary $\chi_u$:
\be \omega\chi_e^I=\log\left|
\chi_u^I \omega' + 1
\over
1-\chi_u^I (\omega - \omega')
 \right|\ee
and the real part goes from 0 to $\omega \chi_e^R=\pi$ if
$(\omega-\omega')\chi_u^I$ crosses 1.
This relation is clearly discontinuous if $\omega\neq\omega'$. We
illustrate this in Fig.~5 in the case $\omega=1$ (eikonal) and
$\omega'=1/2$ ($U$ matrix).
 At low energy, the phases are approximately the same.
But at high energy when  $\chi_u^I (s,b)\rightarrow 2$,
$\chi_e$ has a discontinuity: its imaginary part goes to infinity,
and its real part jumps by $i\pi$. On the other hand, the extended
$U$ matrix does not lead to such a singularity if $\omega'=1$.
\begin{figure}
\epsfysize=45mm
\epsfbox{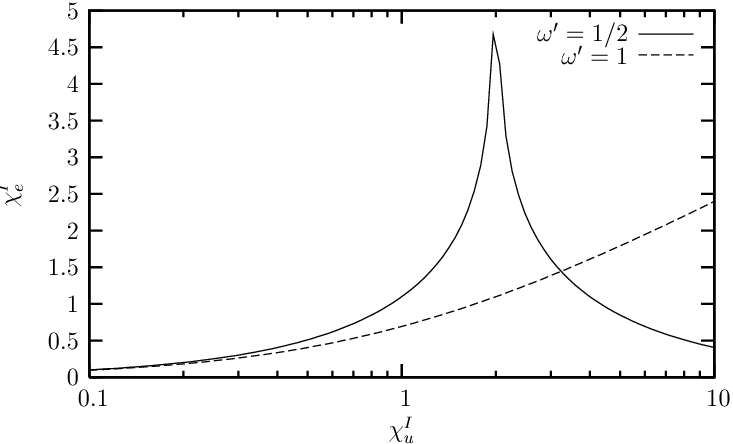}
\caption{The relation between the Born terms of the U-matrix
scheme and of the eikonal scheme, in the case of a purely imaginary
$\chi_u$. The real part of $\omega\chi_e$ is discontinuous
and goes from 0 to $i\pi$ at $\chi_u=2$.
 }
\end{figure}
\section{Non-linear equations}
All previous schemes can be recast as non-linear equations,
which may be reminiscent of those obtained in QCD from gluon
saturation.

The simplest way \cite{CS-nl04} to get these is first to take the
derivative
of $G$ with respect to $\chi$ in (\ref{exteik}) and (\ref{extumat}):
\be
{dG\over d\chi}=1+i\omega G
\ee
in the eikonal case,
and
\be
{dG\over d\chi}=(1+i\omega' G)^2
\ee
for the $U$ matrix.
To make a tentative connection with saturation, we shall consider
a purely imaginary Born term, and we shall write $G=ig$. Assuming
$\chi=ig_0 x^{-\Delta}$, we can write the above equations as an evolution
in $y=\log(1/x)$ at fixed $\bb$:
 \begin{eqnarray}
 \frac{dg}{dy} = {\Delta\over\omega} \log(1-\omega g) (1-\omega g)\,. \label{nl-eik}
 \end{eqnarray}
in the eikonal case
and
 \begin{eqnarray}
 \frac{dg}{dy} =\Delta g  (1-\omega' g) \label{nl-um1}
 \end{eqnarray}
for the $U$ matrix.

We see that the $U$ matrix schemes lead to equations which look
more natural than the corresponding ones in the eikonal case,
as it is hard to imagine how saturation would lead to a log containing the amplitude.

 One can further generalize these equation to reproduce
 Eqs. (\ref{2cl}) and (\ref{1cl}).
    The corresponding nonlinear equation will be
 \begin{eqnarray}
 \frac{d g}{dy} = {\gamma\Delta\over\omega}
  (1- (1 - g)^{1/\gamma})  (1- \omega g)\,. \label{nl-2cl}
 \end{eqnarray}
  It is easily seen that when  $\gamma=1$ we obtain the  non-linear
equation for the $U$ matrix.
  In the case $\omega=1$ Eq.(28) amounts to  the standard logistic equation,
  and leads to the extended $U$-matrix unitarization scheme.

\section{Conclusion}

In this paper we presented two new unitarization schemes which generalize the usual eikonal and $U$
matrix unitarization schemes. We showed that
they belong to two wide classes which cannot be mapped analytically one onto the other.
We showed however that it is possible to build a more general scheme which interpolates
between the two.

The basic behavior of $G(s,\bb)$ as a function of $s$ is mostly constrained by the value of
$\omega$ (or $\omega'$), but not by the details of the unitarizing map. To illustrate this
point, we show in Fig.~6 the behavior of $G(s,\bb=0)$ in our interpolating scheme (\ref{1cl}),
for $\gamma=9$ (close to an eikonal) and for $\gamma=1$ (U matrix), for $\omega=1$ or $1/2$.
Clearly, the large-$s$ behavior of the amplitude is controlled by $\omega$, and not by
$\gamma$.
\begin{figure}[!ht]
\epsfxsize=75mm
\epsfbox{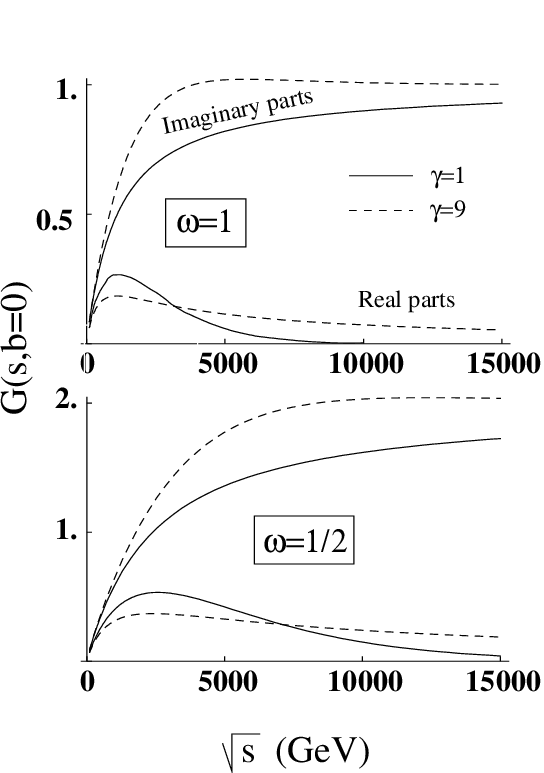}
\caption{The behavior of $G(s,\bb=0)$ from (\ref{1cl}) for various choices of $\gamma$ and $\omega$.
 }
\end{figure}

So the question of the asymptotic behavior of the elastic amplitude
remains open.
It is possible to build an infinite number of schemes in which the amplitude will saturate
at the black disk limit, but there also exists an infinite number of schemes
in which it will exceed
it and eventually converge to the full unitarity limit.

Up to now we do not have a  decisive argument to choose one class of unitarization
over the other. One possibility would be to fit the existing data to determine $\gamma$
and $\omega$. It is however known that this is possible in eikonal schemes, so that it
is unlikely that the constraints will be very stringent. However, the prediction for
the total cross sections in these two classes of unitarization
have large differences (see, for example \cite{physlett,TTst}) for the LHC energy region,
and hence we may know soon which is realized.

\acknowledgments{ The authors would like to thank
  J. Fischer and E. Martynov for helpful discussions.
 O.S. gratefully acknowledges financial support
  from FRNS and INFN and would like to thank the University of Li\`{e}ge
  and Torino University for the hospitality.
    }


\begin{thebibliography}{9}
\bibitem{Predazzi66} E. Predazzi, Ann. of Phys.(N.Y.), {\bf 36},
228; {\bf 36}, 250 (1966).
%
\bibitem{dif04}
J.~R.~Cudell and O.~V.~Selyugin,
  Czech.\ J.\ Phys.\  {\bf 54}, A441 (2004)
  [arXiv:hep-ph/0309194].
%
\bibitem{Shiff} L.I. Shiff, Phys. Rev. {\bf 103} 443 (1956).
%
\bibitem{ChengWu}
H.~Cheng and T.~T.~Wu,
  Phys.\ Rev.\ Lett.\  {\bf 22}, 666 (1969).
%
\bibitem{Blokhintzev} D.I. Blochinzev, Nuovo Cimento {\bf 30}, 1094 (1963).
%
\bibitem{Barbashev} B.M. Barbashov, D.I. Blochinzev, V.V. Nesterenko, V,I. Pervushin,
Fiz. Elem. Chastits At. Yadra (PEPAN) {\bf 4}, 623 (1973).
%
\bibitem{Goldberger} M.L. Goldberger, and K.M. Watson, {\it Collision Theory}
(John Wiley \& Sons, N.Y., (1964).
%
\bibitem{Gotsman}
E.~Gotsman, E.~M.~Levin and U.~Maor,
  Phys.\ Rev.\  D {\bf 49}, R4321 (1994)
  [arXiv:hep-ph/9310257].
%
\bibitem{bg}
R.~Blankenbecler and M.~L.~Goldberger,
  Phys.\ Rev.\  {\bf 126} 766 (1962).
%
 \bibitem{Logunov}
  A.~A.~Logunov, V.~I.~Savrin, N.~E.~Tyurin and O.~A.~Khrustalev,
  Teor.\ Mat.\ Fiz.\  {\bf 6}, 157 (1971).
%
 \bibitem{Chrustalev}   V.~I.~Savrin, N.~E.~Tyurin and O.~A.~Khrustalev,
  Fiz.\ Elem.\ Chast.\ Atom.\ Yadra (PEPAN) {\bf 7}, 21 (1976).
%
 \bibitem{Martiros}  K.~A.~Ter-Martirosyan,
  Pisma Zh.\ Eksp.\ Teor.\ Fiz.\  {\bf 15}, 734 (1972).
%
\bibitem{Kaidalov1}   A.~B.~Kaidalov, L.~A.~Ponomarev and K.~A.~Ter-Martirosian,
  Yad.\ Fiz.\  {\bf 44}, 722 (1986)
  [Sov.\ J.\ Nucl.\ Phys.\  {\bf 44}, 468 (1986)].
%
\bibitem{Martynov1}
  M.~Giffon, E.~Martynov and E.~Predazzi,
  Z.\ Phys.\  C {\bf 76}, 155 (1997).
%
\bibitem{physlett}  J.~R.~Cudell and O.~V.~Selyugin,
  Phys.\ Lett.\  B {\bf 662}, 417 (2008)
  [arXiv:hep-ph/0612046].
%
 \bibitem{TT1}   S.~M.~Troshin and N.~E.~Tyurin,
  Phys.\ Lett.\  B {\bf 316}, 175 (1993)
  [arXiv:hep-ph/9307250].
%
  \bibitem{TT07}   S.~M.~Troshin and N.~E.~Tyurin,
  Int.\ J.\ Mod.\ Phys.\  A {\bf 22}, 4437 (2007)
  [arXiv:hep-ph/0701241].
%
\bibitem{Miettinen}  H.~I.~Miettinen and G.~H.~Thomas,
  Nucl.\ Phys.\  B {\bf 166}, 365 (1980).
%
\bibitem{CS-nl04}
  O.~V.~Selyugin and J.~R.~Cudell,
  arXiv:hep-ph/0408129,
presented at the 11th International Conference on Symmetry Methods in Physics (SYMPHYS-11), Prague, Czech Republic, 21-24 Jun 2004.
%
\bibitem{TTst}   V.~A.~Petrov, A.~V.~Prokudin, S.~M.~Troshin and N.~E.~Tyurin,
  J.\ Phys.\ G {\bf 27}, 2225 (2001).
\end{thebibliography}
\end{document}